\documentclass[twocolumn,showpacs,superscriptaddress,10pt]{revtex4-1}
\usepackage{amsmath,amssymb,graphicx}
\begin{document}

\title{Dichroism for Orbital Angular Momentum using\\
Stimulated Parametric Down Conversion}
\author{J. Lowney}
\affiliation{College of Optical Sciences, University of Arizona, Tucson, AZ 85721-0094, USA} 
\author{T. Roger}
\affiliation{Institute of Photonics and Quantum Sciences, School of Engineering and Physical Sciences, Heriot-Watt University, EH14 4AS Edinburgh, UK} 
\author{D. Faccio}
\affiliation{Institute of Photonics and Quantum Sciences, School of Engineering and Physical Sciences, Heriot-Watt University, EH14 4AS Edinburgh, UK} 
\affiliation{College of Optical Sciences, University of Arizona, Tucson, AZ 85721-0094, USA}
\author{E. M. Wright}\email{Corresponding author: ewan@optics.arizona.edu}
\affiliation{College of Optical Sciences, University of Arizona, Tucson, AZ 85721-0094, USA} 
%
%
\begin{abstract}
We theoretically analyze stimulated parametric down conversion as a means to produce dichroism based on the orbital angular momentum (OAM) of an incident signal field.  The nonlinear interaction is shown to provide differential gain between signal states of differing OAM, the peak gain occurring at half the OAM of the pump field.
\end{abstract}

\maketitle
\section{Introduction}
One usage of the term dichroism in optics is to describe the differential loss of monochromatic light in one of two orthogonal polarization states with respect to some reference axis.  This definition encompasses the case of both linear and circular dichroism, and also gain and/or loss if one allows for negative absorption.  For example, circular dichroism refers to case in which right-handed (RHC) and left-handed circular (LHC) polarizations experience different propagation losses in the dichroic medium.  As is well known the RHC and LHC polarized states represent two orthogonal spin angular momentum (SAM) states for the light field.  Then another way to express circular dichroism is that it is a dichroism based on the spin angular momentum (SAM) state of the incident light field, or simply dichroism for SAM.

The goal of this paper is propose and theoretically investigate stimulated Parametric Down Conversion (PDC) as a means to produce dichroism based on the orbital angular momentum (OAM) of an incident signal field.  The nonlinear interaction is shown to provide differential gain between signal states of differing OAM, the peak gain occurring at half the OAM of the pump field. Stimulated PDC involving fundamental and second-harmonic fields carrying OAM has previously been explored both experimentally and theoretically but mainly in the context of the conservation of OAM for the PDC process \cite{CaeAlmSou02,HugMarCae06,DevPas07} as opposed to creating dichroism for OAM.  A recent paper has discussed circular dichroism that has its origin in the OAM of a beam incident on a non-chiral structure \cite{ZamVidMol14} whereas here we elucidate a means to produce dichroism that acts on the incident beam OAM directly.

The remainder of this paper is organized as follows.  Section \ref{Sec2} describes the geometry and governing equations for our system, and Sec. \ref{Sec3} presents a simplified analytic theory of stimulated PDC and dichroism for OAM for pump and signal beams that are perfect optical vortices.  In Sec. \ref{Sec4} numerical results are presented for the case of signal and pump fields that are imperfect optical vortices and also signal fields that are Laguerre-Gaussian beams.  Specifically, we demonstrate that stimulated PDC can be used to create gain for a band of OAM states of an incident signal beam with absorption outside this band.  Finally summary and conclusions are given in Sec. \ref{Sec5}
\section{Basic geometry and equations}\label{Sec2}
Our basic model involves propagation in the transparency region of a uniaxial nonlinear optical crystal.  More specifically we consider propagation along a principal axis to avoid the effects of beam walk-off and assume type I phase-matching conditions.  In our model of stimulated PDC a signal field at the fundamental frequency $\omega_1$ is incident on the crystal along with a pump field at the second-harmonic (SH) frequency $\omega_2=2\omega_1$.  For the type I phase-matching assumed the signal field is an ordinary wave of refractive-index $n_1$ and the pump is an extraordinary wave with refractive-index $n_2$.  Then, choosing the z-axis as the propagation direction, denoting the complex slowly-varying field amplitudes of the fundamental and SH fields as $A_1(x,y,z)$ and $A_2(x,y,z)$, and following the derivation and notation of Ref. \cite{Boyd}, we obtain the following paraxial wave equations for the signal $(j=1)$ and pump $(j=2)$ fields
\begin{eqnarray}
\frac{\partial A_1}{\partial z}&=&{i\over 2k_1} \nabla_\perp^2 A_1 + \frac{2i\omega_1^2d_{eff}}{k_1c^2}A_2A_{1}^*e^{-i\Delta k z}  ,\nonumber \\
\frac{\partial A_2}{\partial z}&=&{i\over 2k_2} \nabla_\perp^2 A_2 + \frac{i\omega_2^2d_{eff}}{k_2c^2}A_1^2 e^{i\Delta k z}  ,
\end{eqnarray}
where $\nabla_\perp^2={\partial^2\over \partial x^2}+{\partial^2\over \partial y^2}$ is the transverse Laplacian describing beam diffraction, $d_{eff}$ is the effective nonlinear coefficient,  $k_j=n_j\omega_j/c$ gives the z-component of the wavevector for the two fields, and $\Delta k=2k_1-k_2$ is the wavevector mismatch.  Throughout this paper we assume the case of non-critical phase-matching in a LBO crystal and a fundamental wavelength of $\lambda_1=1.064$ $\mu$m for which $n_1=n_2=n=1.6$, $\Delta k=0$, and $d_{eff}=0.83$ pm/V.  Then introducing the parameter $\eta=(2\omega_1d_{eff}/n_1c)$ the propagation equations may be written as
\begin{eqnarray}\label{Aeqs}
\frac{\partial A_1}{\partial z}&=&{i\over 2k_1} \nabla_\perp^2 A_1 + i\eta A_2A_{1}^*  ,\nonumber \\
\frac{\partial A_2}{\partial z}&=&{i\over 4k_1} \nabla_\perp^2 A_2 + i\eta A_1^2  .
\end{eqnarray}
These propagation equations for stimulated PDC are to be solved for input fields that have cylindrically symmetric intensity profiles and carry OAM specified by the winding numbers $m_1$ for the signal and $m_2$ for the pump
\begin{equation}\label{initU}
A_j(x,y,z=0)=\alpha_j U_j(\rho,z=0)e^{im_j\phi}, \quad j=1,2
\end{equation}
where $(\rho,\phi)$ are the transverse coordinates in cylindrical coordinates, and the complex coefficients $\alpha_j$ are used to control the input powers of the fundamental and second-harmonic fields along with the relative phase $\theta$ between the input fundamental and SH fields.  Here $U_j(x,y,z)\equiv U_{j}(\rho,z)$ are normalized cylindrically symmetric field profiles which describe the input fields at $z=0$ and their linear propagation to the output at $z=L$.  The output powers in the fundamental and SH fields can be expressed as
\begin{equation}
P_j(L) = {1\over 2}\epsilon_0  n c \int_{-\infty}^\infty dx \int_{-\infty}^\infty dy |A_j(x,y,L)|^2, \quad j=1,2  .
\end{equation}
We furthermore define the output signal power
\begin{eqnarray}
P_s(L) =&\nonumber \\
 {1\over 2}\epsilon_0  n c & \left | \int_{-\infty}^\infty dx \int_{-\infty}^\infty dy~U_1^*(x,y,L)A_1(x,y,L) \right |^2
\end{eqnarray}
which represents the power contained in the fundamental field projected onto the normalized input signal mode $U_1(x,y,L)$ evaluated at the output.  In the following we shall examine the net gain for the fundamental field
\begin{equation}\label{G}
G = {P_1(L)\over P_s}  ,
\end{equation}
$P_s=P_1(0)$ being the input power, and the signal gain
\begin{equation}\label{Gs}
G_s = {P_s(L)\over P_s}  .
\end{equation}
In general $P_1(z)>P_s(z)$ and $G>G_s$ since the nonlinear interaction will generate modes in the fundamental field beyond the incident signal mode.  We shall always choose the input pump power somewhat larger than the signal power to avoid excessive pump depletion.
\section{Simplified analytic theory}\label{Sec3}
To set the stage for our numerical simulations we first present a simplified analytic theory of stimulated PDC with OAM and associated dichroism.  In particular, we consider the case that both the signal and pump beams are perfect optical vortices (POVs) \cite{OstRicArr13,CheMalAri13}.  A perfect optical vortex of winding number $m$ has a narrow ring intensity profile with an azimuthal phase-twist of $2\pi m$ in the transverse plane of the field.  The key to using POVs is that the ring radius $R$ should be independent of winding number and the same for all interacting fields.  This choice maximizes the spatial overlap of the interacting fields and allows for a treatment that removes issues related to the radial profile of the fields while retaining the azimuthal variation.
\subsection{Perfect optical vortices}
We first present a representation of a POV with frequency $\omega=2\pi c/\lambda$ propagating in a medium of refractive-index $n$.  The POV has a ring shaped intensity profile of radius $R$ and width $W$, $R>>W>>\lambda$, along with a helical phase-front of winding number $m$. (In the ideal case the ring width $W$ would be zero \cite{OstRicArr13}.) We assume that the width $W$ of the POV is sufficiently narrow compared to the ring radius that we may evaluate the properties of the beam around the peak of the ring.  Then for a POV with azimuthal variation $e^{im\phi}$ propagating along the z-axis, the corresponding spiraling wavevector may be written as \cite{RogHeiWri13}
\begin{eqnarray}\label{K}
\vec{K} &=& K_x\vec{e}_x + K_y\vec{e}_y + K_z\vec{e}_z  \nonumber\\
&=&  {m\over R}\cos(\phi)\vec{e}_x + {m\over R}\sin(\phi)\vec{e}_y + K_z\vec{e}_z  ,
\end{eqnarray}
with $R>>\lambda$ the ring radius.  By demanding that $K=k=2\pi nc/\/\lambda$ we obtain for a forward propagating field
\begin{equation}\label{Kz}
K_z = \sqrt{k^2-{m^2\over R^2}}
\approx  k - {1\over 2k}{m^2\over R^2}  ,
\end{equation}
so we get the expected reduction in the z-component of the wavevector due to the skewing associated with the helical phase-front of the POV \cite{DhoSimPad96}.

Based on the above results the slowly varying electric field envelope for a POV evaluated around the peak of the ring may be written as
\begin{equation}
A(\rho=R,\phi,z) = a(z)e^{im\phi}e^{-{iz\over 2k}{m^2\over R^2}}  .
\end{equation}
The utility of this solution rests on the Rayleigh range $z_R=kW^2/2$ being much larger than the medium length $L$ so that the ring width will vary little under propagation through the medium.  Within this approximation the transverse Laplacian in cylindrical coordinates becomes $\nabla_\perp^2\rightarrow {1\over R^2}{\partial^2\over \partial\phi^2}$, thereby neglecting radial expansion of the ring.
\subsection{Stimulated parametric down conversion}
For this development we assume that the pump $(j=2)$ field is much stronger than the signal $(j=1)$ field.  Then the stimulated PDC process, which produces one signal and one idler photon from one pump photon, generates an idler field $(j=3)$ that has winding number $m_3=m_2-m_1$.  Assuming all fields are described by POVs we then write the slowly varying electric fields for the fundamental and second harmonic fields, with $\rho=R$, as
\begin{eqnarray}
A_1(\phi,z) &=& a_1(z)e^{im_1\phi}e^{-{iz\over 2k_1}{m_1^2\over R^2}}+a_3(z)e^{im_3\phi}e^{-{iz\over 2k_1}{m_3^2\over R^2}}, \nonumber \\
A_2(\phi,z) &=& a_2e^{im_2\phi}e^{-{iz\over 2k_2}{m_2^2\over R^2}},
\end{eqnarray}
with $a_2$ independent of $z$ in the undepleted pump beam approximation, and $a_3(0)=0$ with no idler present at the input.  Substituting these fields into Eqs. (\ref{Aeqs}) and using the results from the previous subsection yields the linearized signal-idler equations \cite{Boyd}
\begin{equation}
{da_1\over dz}=i(\eta a_2) a_3^*e^{i\kappa z}, \quad
{da_3\over dz}= i(\eta a_2) a_1^*e^{i\kappa z}  ,
\end{equation}
where the OAM dependent wavevector mismatch for the process is

\begin{equation}\label{kappa}
\kappa = {(m_1-m_2/2)^2\over k_1R^2} ,
\end{equation}
and we note that phase-matching $\kappa=0$ requires $m_1=m_2/2$.  These equations may be solved for the fields at the output of the crystal of length $L$ \cite{Boyd}
\begin{eqnarray}
a_1(L) &=& a_1(0)\left ( \cosh(gL) - {i\kappa\over g}\sinh(gL)\right ), \nonumber \\
a_3(L) &=& a_1^*(0)\left ({\kappa\over g}\right )\sinh(gL) ,
\end{eqnarray}
where $g=\sqrt{\eta^2|a_2|^2-\kappa^2/4}$ is the growth rate if the argument of the square root is positive. The field intensities are given by $I_j(z)={1\over 2}\epsilon_o nc|a_j(z)|^2$ in terms of which the growth rate may be written as
\begin{equation}\label{g}
g = \sqrt{\beta I_p -\kappa^2} ,
\end{equation}
with $I_p=I_2(0)$ is the pump intensity at the peak of the ring and $\beta=(8\omega_1^2d_{eff}^2/\epsilon_0n^3c^3)$.  Using this solution the signal gain may be expressed as
\begin{equation}
G_s=\frac{I_1(L)}{I_1(0)}=\left | \cosh(gL) - {i\kappa\over g}\sinh(gL)\right |^2  .
\end{equation}
Note that under phase-matching $\kappa=0$ the peak signal gain is
\begin{equation}\label{Gpeak}
G_{peak}=\cosh^2(\sqrt{\beta I_p}),
\end{equation}
which increases with pump intensity.

In summary, the simplified analytic solution demonstrates that phase-matching for the stimulated PDC process depends on the following combination of the winding numbers of the signal and pump beams
\begin{equation}\label{Delta}
\Delta = m_1-{m_2\over 2} ,
\end{equation}
whereas the peak signal gain varies with the input intensity.
\subsection{Dichroism for OAM}
\begin{figure}
\begin{center}
\includegraphics*[width=9cm]{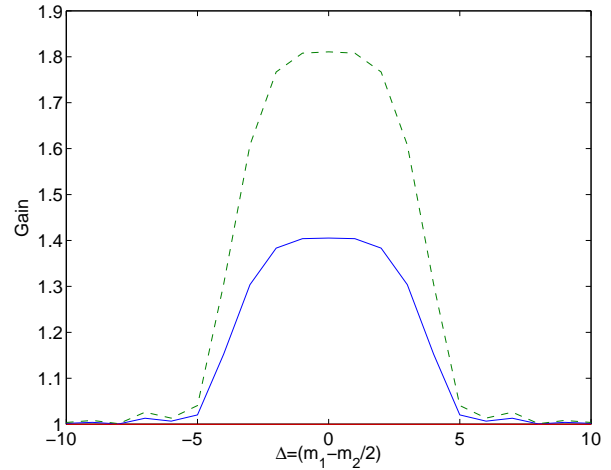} \caption{Signal gain $G_s$ (solid line) and net gain (dashed line) versus OAM difference $(m_1-m_2/2)$ for a LBO crystal of length $L=2$ mm, pump intensity $I_2=0.5$ GW/cm$^2$, and ring radius $R=35~\mu$m.  The discrete data points are connected by a solid line as a visual aid.} \label{Fig1}
\end{center}
\end{figure}

Figure \ref{Fig1} shows an illustrative example of the signal gain $G_s$ (solid line) and net gain $G$ (dashed line) versus OAM difference $\Delta=(m_1-m_2/2)$ for a LBO crystal of length $L=2$ mm, pump intensity $I_p=0.5$ GW/cm$^2$, and a ring radius $R=35~\mu$m. This figure reveals that significant gain occurs for a limited range of OAM values centered on $\Delta=0$, that is around $m_1=m_2/2$.  The fact that the peak of the net gain defined in Eq.~(\ref{G}), which includes both the signal and idler, exceeds the peak of the signal gain defined in Eq. (\ref{Gs}) reflects the fact that a significant idler intensity is generated in this example, but we note that gain appears over a similar range in both cases. The full-width for the parametric gain profile may be estimated by requiring $\kappa L=\pi$ at the edges for the phase-mismatch to diminish the gain, which yields
\begin{equation}\label{dm}
\delta m = 2\sqrt{{\pi k_1R^2\over L}}
\end{equation}
This full-width scales as $\delta m\propto R/\sqrt{\lambda_1 L}$ analogous to the spiral bandwidth used in spontaneous PDC if we replace the Gaussian waist of the pump beam with the ring radius \cite{MiaYaoBar11, McClaren, Alarcon}.  For this reason we refer to $\delta m$ as the spiral bandwidth. For the chosen parameters this yields a spiral bandwidth of $\delta m=9$ in reasonable agreement with Fig. \ref{Fig1}. Note also that the spiral bandwidth is independent of the winding numbers of the incident fields.

The stimulated PDC process therefore provides differential gain between different OAM states of the signal beam, and in this sense acts as a dichroic element based on the signal OAM with the peak gain centered at $m_1=m_2/2$ and spiral bandwidth given by Eq. (\ref{dm}). Furthermore, if we choose $|m_2|>\delta m$ then for $m_2>0$ we can create the situation such that only OAM states with $m_1>0$ experience significant gain, and vice versa for $m_2<0$.

This concludes our discussion of the simplified analytic theory.  Next we turn to numerical simulations using more realistic and practical beam profiles that will expose more general features of stimulated PDC and associated dichroism for OAM.
\section{Numerical simulations}\label{Sec4}
In this Section we present numerical simulations of stimulated PDC for more realistic types of input beams.  The simulations are based on Eqs. (\ref{Aeqs}) with initial conditions corresponding to signal and pump beams carrying OAM as in Eqs. (\ref{initU}). A standard beam propagation method is employed for the nonlinear propagation \cite{FeiFle80}.
\subsection{Imperfect optical vortices}
Here we consider input beams that have a ring structure plus helical phase-fronts but they are not ideal POVs so we term them imperfect optical vortices (IOVs).  In particular, with reference to Eqs. (\ref{initU}) we write the radial profiles of the input fields as
\begin{equation}\label{Uj}
U_j(\rho,z=0) = {\cal N}_j\cdot \rho^{m_r}e^{-\rho^2/w_0^2}, \quad j=1,2
\end{equation}
where ${\cal N}_j$ are normalization constants, $w_0$ a Gaussian beam waist, and $m_r$ is a positive integer. Equations (\ref{Uj}) describe annular shaped beams of ring radius $R=w_0\sqrt{m_r/2}$, and for our numerics we choose $w_0=15~\mu$m in which case the ring radius is $R=35~\mu$m for $m_r=11$.  We note that these initial conditions do not coincide with the familiar Laguerre-Gaussian modes of free-space unless $|m_j|=m_r$ so that these IOVs will generally change their functional form under linear propagation. For our parameters the fundamental Rayleigh range is $z_R\simeq 1$ mm whereas the medium length is $L=2$ mm, so the IOVs experience non-negligible diffraction over the medium length.

\begin{figure}
\begin{center}
\includegraphics*[width=9cm]{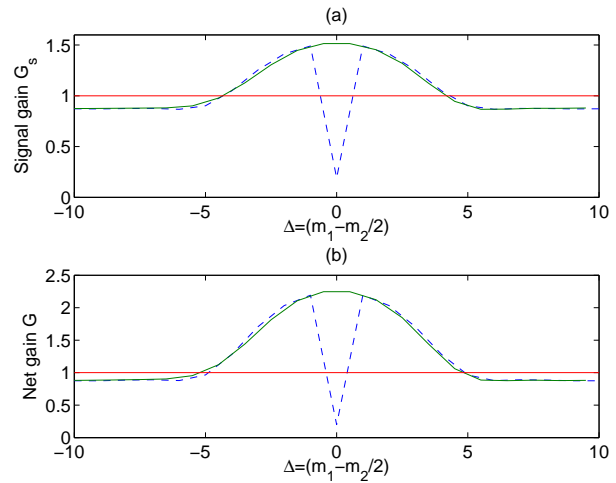} \caption{(a) Signal gain $G_s$ versus OAM difference $\Delta=(m_1-m_2/2)$ for $m_2=0$ (dotted line) and $m_2=11$ (solid line), and (b) same for the net gain G. Parameters used are for LBO and $m_r=11, w_0=15~\mu$m giving $R=35~\mu$m, an input pump power of $P_2(0)=66$ kW, and a signal power of $P_s=0.2P_2(0)=13$ kW, giving a peak intensity $I_p=2.2$ GW/cm$^2$. The discrete data points are connected by a solid line as a visual aid.} \label{Fig2}
\end{center}
\end{figure}

Figure \ref{Fig2} shows illustrative examples of stimulated PDC using IOVs with parameters $m_r=11, w_0=15~\mu$m giving $R=35~\mu$m, an input pump power of $P_2(0)=66$ kW, and a signal power of $P_s=0.2P_2(0)=13$ kW, these parameters yielding an intensity of $I_p=2.2$ GW/cm$^2$ around the peak of the pump beam.  The signal gain $G_s$ given in Eq. (\ref{Gs}) is plotted as a function of OAM difference $\Delta=(m_1-m_2/2)$ for the cases with $m_2=0$ (dotted line) and $m_2=11$ (solid line).

The results in Fig. \ref{Fig2}(a) display qualitative similarities and differences with the simplified analytic theory in Figure \ref{Fig1} that we now discuss.  First, except for $\Delta=0$ the results for the two different pump winding numbers $m_2=0,11$ agree very well, this being expected from the simplified theory.  However, for the case of zero winding number for both the pump and probe $m_1=m_2=\Delta=0$ (dashed line) the signal gain shows an absorption dip.  This arises since under this condition there is a resonant interaction between the injected fundamental field and the SH field which preserves the winding number of each field, and which depends on the relative phase $\theta$ between the signal and SH.  For the case shown $\theta=\pi/4$ this yields absorption, whereas for $\theta=-\pi/4$ signal gain occurs \cite{DevPas07}.  In contrast the case with $m_2=11$ shows no such absorption at $\Delta=0$.  This is because the resonant interaction between the fundamental and SH fields at $\Delta=0$ requires $m_1=m_2/2$ which cannot be satisfied for integer $m_1$ if $m_2$ is odd, but the absorption dip does appear at $\Delta=0$ if $m_2$ is even.  So excluding the dip at $\Delta=0$ the results for $m_2=0,11$ agree well. Note also that in Fig. \ref{Fig2}(a) the signal gain turns to absorption for larger values of $|\Delta|$.  This background absorption arises from conversion of the fundamental field with OAM $m_1$ to SH with OAM $2m_1$ (generally distinct from the input SH with OAM $m_2$.)  The magnitude of this background absorption increases as the input signal power is increased.  Furthermore the width of the central peak in Fig. \ref{Fig2}(a) is around nine which is close to the spiral bandwidth $\delta m=9$ obtained from Eq. (\ref{dm}), the parameters being the same as Fig. \ref{Fig1}.

Figure \ref{Fig2}(b) shows the same as (a) but for the net gain given by the total fundamental output power divided by the input signal power.  Plots (a) and (b) show the same features but the net gains are larger than the signal gains due to the inclusion of the idler power in the net gain.  The reason for this figure is to demonstrate that the common features appear in both gains and it is simpler to measure the net gain experimentally than to isolate the signal gain.  Both gain measurements would demonstrate that the differential gain or loss between different signal OAM states depends on the OAM difference $\Delta=(m_1-m_2/2)$.

\begin{figure}
\begin{center}
\includegraphics*[width=9cm]{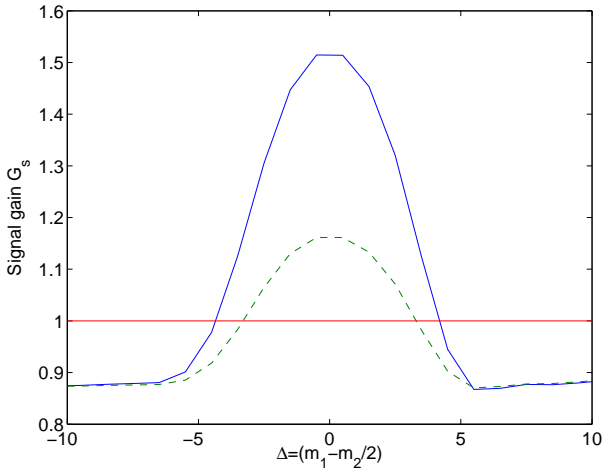} \caption{Signal gain $G_s$ versus the OAM difference $\Delta=(m_1-m_2/2)$ for the same parameters as Fig. \ref{Fig2}(a) with $m_2=11$, and a pump intensity of $I_p=2.2$ GW/cm$^2$ (solid line) and $I_p=1.1$ GW/cm$^2$ (dashed line), the signal power being held constant at $P_s=0.2P_2(0)=13$ kW. The discrete data points are connected by a solid line as a visual aid.} \label{Fig3}
\end{center}
\end{figure}

Further features of the signal gain $G_s$ are illustrated in Fig. \ref{Fig3}.  This figure shows the signal gain versus the OAM difference $\Delta=(m_1-m_2/2)$ for the same parameters as Fig. \ref{Fig2}(a) with $m_2=11$, and a pump intensity of $I_p=2.2$ GW/cm$^2$ (solid line) and $I_p=1.1$ GW/cm$^2$ (dashed line), the signal power being held constant at $P_s=13$ kW.  As expected on the basis of Eq. (\ref{Gpeak}) the peak signal gain increases with pump intensity, and given a peak signal gain of $G_{peak}=1.5$ for the higher pump intensity Eq. (\ref{Gpeak}) predicts $G_{peak}=1.23$ for the lower pump intensity, in reasonable agreement with the numerics (recall that the simple theory does not account for the background absorption due to second harmonic generation (SHG) that is present in the numerics).  In contrast we see that the background absorption is the same in both cases.  This follows since the background absorption arises from SHG of the fundamental field of winding number $m_1$ to create a SH field with winding number $2m_1$, distinct from the pump SH field with winding number $m_2=11$, and this depends dominantly on the signal properties alone, not the pump properties.

\begin{figure}
\begin{center}
\includegraphics*[width=9cm]{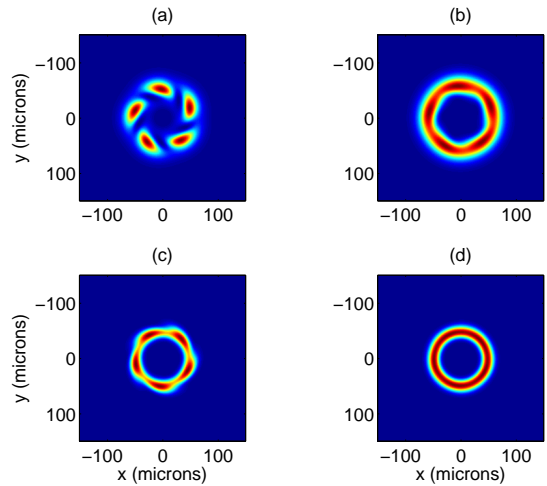} \caption{Examples of the fundamental (top row) and SH (bottom row) output transverse intensity profiles for two different values of the pump power $P_2(0)=66$ kW (left column) and $P_2(0)=6.6$ kW (right column), with $P_s=0.2P_2(0)$, all other parameters being the same as in Fig. \ref{Fig2}. The winding numbers of the fundamental and SH fields were chosen as $m_1=8,m_2=11$ so that $\Delta=2.5$, and the generated idler will have winding number $m_3=m_2-m_1=5$.} \label{Fig4}
\end{center}
\end{figure}

Figure \ref{Fig4} shows illustrative examples of the fundamental (top row) and SH (bottom row) output transverse intensity profiles for two different values of the pump power $P_2(0)=6.6,66$ kW, with $P_s=0.2P_2(0)$, all other parameters being the same as in Fig. \ref{Fig2}. The winding numbers of the fundamental and SH fields were chosen as $m_1=8,m_2=11$ so that $\Delta=2.5$, and the generated idler will have winding number $m_3=m_2-m_1=3$.  For plots (a,c) the pump power is $P_2(0)=66$ kW, and the fundamental intensity profile in plot (a) shows a five-lobe structure which arises from azimuthal beating between the signal and idler fields with azimuthal periodicity $2\pi/|m_1-m_3|=2\pi/5$.  The pronounced lobes reflect the fact that a strong idler is generated in this case (as also evidenced by the difference between the net gain and signal gain in Fig. \ref{Fig2} for $\Delta=2.5$).  The five lobe structure is also evident but to a lesser degree in the corresponding intensity profile for the SH shown in plot (c).  Plots (b,d) show the same thing for a pump power $P_2(0)=6.6$ kW, the key difference being that a weaker idler is generated and the five-lobe structure is less well pronounced.  For even lower pump powers the intensity profiles tend closer to rings.
In summary, many of the features present in the simplified analytic model are also present using signal and pump beams that are IOVs.  The simplified theory did not include the SHG process so it did not account for the resonant SHG that occurs for $m_1=m_2/2$, or the background absorption of the signal due to generation of a SH field at $m_2=2m_1$.  The simplified model did capture the spiral bandwidth of the stimulated PDC process.  It then follows that the dichroism for OAM displayed by the simple model may also be realized using IOVs.  A key distinction is that whereas the simplified analytic theory only shows differential gain between signal OAM states, the full theory with IOVs shows gain for a band of OAM states and loss outside that band, and in this sense the full theory is richer.
\begin{figure}
\begin{center}
\includegraphics*[width=9cm]{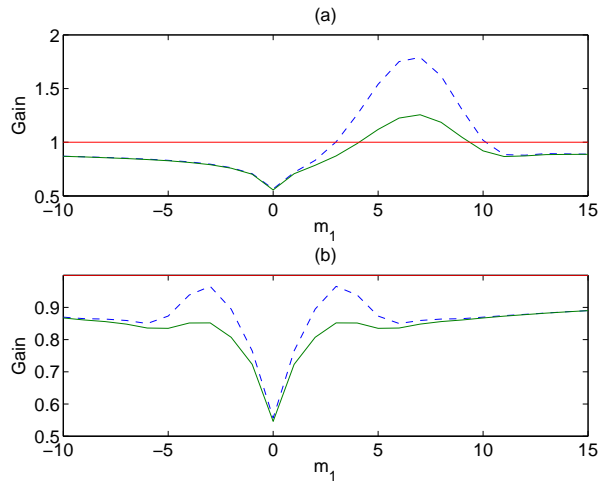} \caption{Plot (a) shows the signal gain $G_s$ (solid line) and net gain $G$ (dashed line) as functions of the signal beam winding number $m_1$ for a pump winding number $m_2=11$, and (b) shows the same for $m_2=0$.  Parameter values are $m_r=11, w_0=15~\mu$m giving $R=35~\mu$m, an input pump power of $P_2(0)=66$ kW, and a signal power of $P_s=0.2P_2(0)=13$ kW.  The discrete data points are connected by a solid line as a visual aid.} \label{Fig5}
\end{center}
\end{figure}
\subsection{Laguerre Gaussian signal}
For our second example we consider the case that the pump beam is an IOV as in Eq. \eqref{Uj} but the signal beam is a Laguerre-Gaussian (LG). For input beams other than POVs or IOVs the spatial overlap of the signal and pump beams introduces new features beyond the simplified theory and we use the LG beams as an illustrative example due to their relative ease of generation in the laboratory.  In particular we consider LG signal modes with radial mode index $p=0$ and winding number $m_1$
\begin{equation}\label{U1}
U_1(\rho,z=0) = {\cal N}_j\cdot \rho^{|m_1|}e^{-\rho^2/w_0^2} ,
\end{equation}
the pump IOV and signal LG being based on the same Gaussian spot size $w_0$.  For $m_1=0$ this is a Gaussian beam peaked on axis whereas for $m_1\ne 0$ this is a ring beam with radius $R_1=w_0\sqrt{|m_1|/2}$, so that the ring radius varies with winding number in contrast to the IOVs.  For a pump beam that is an IOV as in Eq. (\ref{Uj}) the ring sizes of the LG signal and pump beam will coincide when $m_r=|m_1|$.

Figure \ref{Fig5} shows illustrative examples of stimulated PDC using LG signal beams with parameters $m_r=11, w_0=15~\mu$m giving $R=35~\mu$m, an input pump power of $P_2(0)=66$ kW, and a signal power of $P_s=0.2P_2(0)=13$ kW, these parameters yielding an intensity of $I_p=2.2$ GW/cm$^2$ around the peak of the pump beam.  In plot (a) the signal gain $G_s$ (solid line) and net gain $G$ (dashed line) are shown as a function of the signal beam winding number $m_1$ for a pump winding number $m_2=11$.  This figure shows that parametric gain occurs over a band of winding numbers with peak gain centered around $m_1\simeq 7$, with absorption outside of this gain band.  The gain peak is shifted with respect to the phase-matching condition $m_1=m_2/2=5.5$ but this is not surprising since the overlap between the interacting fields, which enters into the strength of the parametric wave interaction, varies with $m_1$.  Although the gain profile is asymmetric the results in plot (a) largely conform to the findings based on the using IOVs.  This example demonstrates that by judicious choice of signal mode structure we can create the dichroism for OAM we elucidated using POVs.

The big difference for LG beams occurs when the winding number of the pump is changed, and is illustrated for $m_2=0$ in Fig. \ref{Fig5}(b) which shows the signal gain $G_s$ (solid line) and net gain $G$ (dashed line) as a function of the signal beam winding number $m_1$.  In contrast to the case of IOVs where changing $m_2$ would simply shift the gain profile along the $m_1$ axis, see Fig. \ref{Fig2}, the gain profiles in plot (b) are distinctly different from those in plot (a).  In particular, for the chosen example the signal field experiences absorption for all values of $m_1$ (we chose the relative phase $\theta=\pi/4$ so that there is absorption at $m_1=0$).  This arises since the phase-matching condition for peak gain now occurs at $m_1=0$ but there is little overlap between the signal and SH fields at that point and therefore little concomitant parametric gain to overcome losses due to second-harmonic generation.  The main observation is that for more general signal and pump beam profiles the gain profile depends on the signal and probe winding numbers independently and not just through the OAM difference $\Delta=(m_1-m_2/2)$.
\section{Summary and conclusions}\label{Sec5}
In summary, we have investigated stimulated Parametric Down Conversion (PDC) as a means to produce dichroism based on the orbital angular momentum (OAM) of an incident signal field.  Specifically, we have demonstrated that stimulated PDC can be used to create gain for a band of OAM states of an incident signal beam with absorption outside this band.  The spiral bandwidth of the gain was shown to depend on beam parameters and the medium length, whereas the peak gain occurs for a signal OAM equal to half that of the pump field, and the peak signal gain increases with the pump power.  This illustrates that stimulated PDC processes can be used to provide significant gain to a particular sign of the probe OAM which could in turn be used, for example, to sculpt the OAM content of an incident signal beam, or bias the oscillating OAM states in an active system such as a laser.  In a similar manner this dichroism could be used to vary the gain for a specific probe OAM dependent on the sign of the OAM of the pump, and this could be used for all-optical switching of the probe.\\

To conclude, we remark that the results in Fig. \ref{Fig5} bear some resemblance to those predicted by Zel'dovich in the early '70s \cite{Zel}.  More specifically, the Zel'dovich effect involves light scattering from an absorbing cylinder.  If the cylinder is not rotating then a probe field incident radially onto it will suffer some absorption. Zel'dovich  showed that if the cylinder is rotating then the probe can experience gain over a range of probe winding numbers, the required energy coming from the energy that needs to be added to sustain the rotation \cite{Zel,Zel2}.  The Zel'dovich effect is therefore another system that can display dichroism for OAM. Another, closely related effect is Penrose superradiance, i.e. amplified scattering waves with angular momentum falling into a rotating black hole \cite{Penrose}.   To elucidate the analogy based on the PDC system, the role of the cylinder is played by the second-harmonic pump and the role of the probe is played by the signal.  In this analogy the probe experiences the refractive-index perturbation induced in the medium by the pump field via the second-order nonlinearity: this perturbation is rotating if $m_2\ne 0$.  More technically, parametric gain around the pump beam ring creates an ergoregion in which energy can be exchanged between fields of differing OAM as dictated by phase-matching.  Then the results in Fig. \ref{Fig5}(b) show that if the cylinder (SH pump) is non-rotating, $m_2=0$, the probe is absorbed for all incident winding numbers, as expected for waves impinging on an absorbing, non rotating cylinder or on a non-rotating black hole.  In contrast when the cylinder (pump beam) is rotating, gain becomes possible.  Stimulated PDC therefore provides a {\emph {nonlinear}} analogue system for the Zel'dovich effect. It is worth noting the similarities and differences of the two systems: in the Zel'dovich effect, loss and gain are described by the same {\emph {linear}} loss coefficient that changes sign depending only on the relative rotation frequencies of the cylinder and probe beam. In the nonlinear PDC system, loss is represented by SHG that funnels energy from the probe into a SH signal that has different OAM with respect to the pump. Gain on the other hand is observed when the correct spatial phase relations are imposed between the pump and probe. OAM PDC dichroism therefore depends on the phase properties of the probe (as in the Zel'dovich effect) and also of the pump. Notwithstanding this difference, the two processes are intriguingly similar, the main point being that both can display dichroism for OAM.
\section{Acknowledgements}
D.F. acknowledges financial support from the European Research Council under the European Unions Seventh Framework Programme (FP/2007-2013)/ERC GA 306559 and EPSRC (UK, Grant EP/J00443X/1).
\end{document}